\documentclass[twocolumn,english]{aastex6}
\usepackage[utf8]{inputenc}
\usepackage{array}
\usepackage{rotfloat}
\usepackage{multirow}
\usepackage{amstext}
\usepackage{amssymb}
\usepackage{graphicx}
\usepackage{wasysym}

\makeatletter

\providecommand{\tabularnewline}{\\}

\usepackage[space]{grffile}
\usepackage{latexsym}
\usepackage{textcomp}
\usepackage{longtable}
\usepackage{tabulary}
\usepackage{booktabs}
\usepackage{array}\usepackage{multirow}\usepackage{amsfonts}
\providecommand{\citet}{\cite}
\providecommand{\citep}{\cite}
\providecommand{\citealt}{\cite}
\newif\iflatexml\latexmlfalse
\DeclareGraphicsExtensions{.pdf,.PDF,.png,.PNG,.jpg,.JPG,.jpeg,.JPEG}

\usepackage[english]{babel}

\shorttitle{The supernova type Ia rate from colliding white dwarfs in triples}
\shortauthors{Toonen et al. }

\makeatother

\usepackage{babel}
\begin{document}

\title{The rate of WD-WD head-on collisions in isolated triples is too low
to explain standard type Ia supernovae }

\begin{abstract}
Type Ia supernovae (Ia-SNe) are thought to arise from the thermonuclear
explosions of white dwarfs (WDs). The progenitors of such explosions
are still highly debated; in particular the conditions leading to
detonations in WDs are not well understood in most of the suggested
progenitor models. Nevertheless, direct head-on collisions of two
WDs were shown to give rise to detonations and produce Ia-SNe - like
explosions, and were suggested as possible progenitors. The rates
of such collisions in dense globular clusters are far below the observed
rates of type Ia SNe, but it was suggested that quasi-secular evolution
of hierarchical triples could produce a high rate of such collisions.
Here we used detailed triple stellar evolution populations synthesis
models coupled with dynamical secular evolution to calculate the rates
of WD-WD collisions in triples and their properties. We explored a
range of models with different realistic initial conditions and derived
the expected SNe total mass, mass-ratio and delay time distributions
for each of the models. We find that the SNe rate from WD-WD collisions
is of the order of $0.1$\% of the observed Ia-SNe rate across all
our models, and the delay-time distribution is almost uniform in time,
and is inconsistent with observations. We conclude that SNe from WD-WD
collisions in isolated triples can at most provide for a small fraction
of Ia-SNe, and can not serve as the main progenitors of such explosions. 
\end{abstract}

\author{S. Toonen}
\affil{Anton Pannekoek Institute for Astronomy, University
of Amsterdam, 1090 GE Amsterdam, The Netherlands } 

\affil{Department of Physics, Technion, Haifa 3200003, Israel}
\author{ H.~B. Perets}
\affil{Department of Physics, Technion, Haifa 3200003, Israel}

\author{A.~S. Hamers}
\affil{Institute for Advanced Study, School of Natural Sciences, Einstein Drive, Princeton, NJ 08540, USA }
\affil{Leiden Observatory, Leiden University, PO Box 9513, NL-2300 RA Leiden, The Netherlands}

\email{* email: toonen@uva.nl}

\section{Introduction}

Type Ia supernovae (Ia-SNe) have played a pivotal role in our understanding
of the structure of the universe and its expansion through their use
as standard candles \citep{phi+93,rei+98,per+99}, as well as the
chemical composition and evolution of galaxies \citep[e.g.][]{pag+97}.
However, despite the significance, the origins of these SNe are still
hotly debated \citep[see e.g.][for reviews]{hil+00,Mao+14}.

It is thought that regular Ia-SNe \citep[there are several classes of peculiar
Ia-SNe, which we do not discuss here, see e.g.][]{Li01, Gra17, Tau17} are powered by the thermonuclear
explosion of a carbon-oxygen (CO) white dwarf (WD), and several astrophysical
scenarios leading to Ia-SNe explosions have been proposed, and one
or even multiple progenitor channels may exist \citep{hil+00,mao+12}.
There are three classical progenitor channels of which two concern
a WD reaching the Chandrasekhar mass limit. This happens either by
accretion from a non-degenerate companion star in the single-degenerate
(SD) channel \citep{Whe+73,nom82}, or by a merger of two CO WDs in
the double-degenerate (DD) channel \citep{Ibe+84,web84}. Another
Chandrasekhar-mass channel, concerns the merger of a WD and an AGB-star
degenerate core  \citep{Kas11, Ilk12, Sok13}. Additionally, sub-Chandrasekhar models
have been considered e.g. through the double-detonation channel after
He-accretion \citet[e.g.][]{woo+86,liv90}. In recent years the DD
scenario has been extended to include WD-WD \textit{collisions} and not
only mergers, but the former were thought to be extremely rare, and
occur only in dense stellar clusters. For this reason they attracted
relatively little attention compared with other WD explosion progenitors.
Such collisions, however, are likely to be observable as type Ia SNe
\citep{Ros+09} and possibly non-standard SNe \citep{Ras+09,Ras+10,Pap+16}.
Recently, it was shown that some triple systems may dynamically evolve
through a quasi-secular process, reminiscent of Kozai-Lidov oscillations
\citep{koz62,Lid62}, but where significant peri-center changes can
occur on a single orbit time-scale \citep{Ant12}, leading to extremely
close peri-center approaches. In particular, it was suggested that
such evolution in triples hosting an inner WD-WD binary could lead
to physical collisions and the production of type Ia SNe \citep{Kat12,Tho11,Kus+13}. 

The WD-WD collision scenario has several advantages. In particular,
the detonation mechanism (shock ignition) is well understood and robust
compared with other progenitor models \citep{Kus+13}, and the model
may provide a range of Ia-SNe with properties consistent with the
observed ones. Physical collisions of WDs have been considered in
the context of dense environments such as globular clusters and the
Galactic centre \citep{Hut85,Sig93}. However, the rates are expected
to be several orders of magnitude below that of the observed Ia-SNe
rate \citep{Ben89}. The rate of collisions arising from a different
channel, namely the evolution of isolated triples, however, was suggested
to be high \citep{Kat12}. Nevertheless, it was never self-consistently
estimated and preliminary calculations suggested it is actually low
\citep{Ham13, Sok+14,Pap+16}. Here we try to close our knowledge gap and
calculate the expected rate of such collisions self-consistently,
test the viability of the isolated-triples WD-WD collision model (hereafter
the WD-collision model) in term of the progenitor production rates,
and derive the delay time distribution (DTD) and collision components
(masses) in such explosions. As we show in the following, we find
that the SNe rate from WD-WD collisions is of the order of at most
$0.1$\% of the observed Ia-SNe rate across all the models we explored,
and the delay-time distribution is almost uniform in time, and is
\emph{inconsistent} with observations and unlikely to explain the
origins of standard Ia-SNe. 

We begin by describing the triple population synthesis we use (section
\ref{sec:met}) and the method applied to couple it with quasi-secular
triple evolution. We then lay out our assumptions and our criteria
for identifying WD-WD collisions in our models and list the range
of initial conditions explored. We then describe our detailed results
for each of our models \ref{sec:Res} and then discuss the results
and summarize in section \ref{sec:concl}. 

\section{Method\label{sec:met}}

We study the collision rate of WDs in triples by simulating the evolution
of populations of triples. The first triple population synthesis studied
were done in the context of destabilized triples due to stellar evolution
\citep{per+12}, which did not account for secular dynamics. Later
\citet{Ham13} and \citet{Nao+16} developed population synthesis
codes which included secular evolution, but do not account for quasi-secular
regime. Here we use a recently developed triple population synthesis
code \texttt{TRES} (\citealt{Too16}; see section \ref{subsec:TRES}
for details) which we complement with a simplified treatment of the
quasi-secular evolution implications for collisions.

The simulation of a triple system starts with three stars on the zero-age
main-sequence in a specific orbital configuration. As the distribution
of masses and orbital parameters of these primordial triples is not
well known, we apply different model populations to assess the systematic
error on our calculations (Sect. \ref{subsec:Primordial-triples}). 

The evolution from the main-sequence on-wards is simulated with the
triple evolution code \texttt{TRES} (Sect. \ref{subsec:TRES}).
We consider triples that avoid mass transfer and remain dynamically
stable throughout their evolution. Our triples evolve into a triple
WD (3WD), or a double WD in the inner binary with a stellar tertiary
(2WD). Generally, the tertiary has a low mass ($m_{3}<0.95{\rm {M_{\odot}}}$)
such that it does not evolve into a WD within a Hubble time. Subsequently,
a collision occurs between the WDs in the inner binary that leads
to the SNIa explosion. The collision can occur due to secular dynamics
in three-body systems which drive the inner eccentricity to high values.
However \citet{Kat12} demonstrated the collision rate may be significantly
enhanced for marginally hierarchical systems due to a breakdown of
the secular approximation in the quasi-secular regime. We apply two
methods to extract the systems that have evolved to a marginal hierarchy
(Sect.\ref{subsec:Marginal-hierarchical-systems}). For these systems,
the tertiary can significantly change the angular momentum of the
inner binary by order unity during a pericenter passage. \citet{Kat12}
show that the angular momentum phase space is stochastically scanned,
such that after a large number of pericenter passages a collision
can be expected. Subsequently, the time to reach a collision $t_{{\rm col}}$
roughly follows a Poissonian distribution with a mean of:
\begin{equation}
\tilde{t}_{{\rm col}}=\frac{a_{{\rm in}}}{4R_{{\rm wd}}}P_{{\rm in}},\label{eq:tcol}
\end{equation}
where $P_{{\rm in}}$ is the period of the inner binary, and $R_{{\rm wd}}$
is the radius of a WD here taken to be $10^{9}$ cm. Throughout our
analysis we assume that a triple entering into the quasi-secular regime
leads to the inner binary direct collision on this given timescale.

\subsection{Primordial triples\label{subsec:Primordial-triples}}

We perform simulations for six sets of primordial triples. These differ
with respect to the distributions of stellar and orbital parameters
(Table \ref{tbl:init_param}). In the standard model STD, we assume
the mass ($m_{1}$) of the initially most massive star in the inner
binary (hereafter primary) follows the Kroupa initial mass function
\citep[IMF,][]{Kro93}. The mass ratios of the inner binary ($q_{{\rm inner}}\equiv m_{2}/m_{1}$,
where $m_{2}$ is the mass of the secondary) are distributed uniformly
\citep{San12,Duc13,Moe17}. We assume that the mass $m_{3}$ of the
outer companion (hereafter tertiary) is uncorrelated to that of the
inner stars. This is consistent with observations of binaries with
wide orbits \citep{Moe17}. Furthermore, in model STD the inner and
outer semi-major axes ($a_{{\rm inner}}$ and $a_{{\rm outer}}$)
are distributed uniformly in log-space ($N\propto1/a$, \citet{Abt83}
between $5{\rm {R}}{}_{\odot}$ and $5\times10^{6}{\rm {R_{\odot}}}$.
The eccentricities $e_{{\rm inner}}$ and $e_{{\rm outer}}$ are distributed
thermally \citep{Heg75} between 0 and 1. The mutual inclination follows
a circular uniform distribution between 0 and $\pi$. We assume that
the arguments of pericenter and the lines of ascending nodes of both
the inner and outer orbits are distributed uniformly between $-\pi$
and $\pi$. We assume the stars have Solar metallicities. Finally,
we adopt a constant binary fraction of 40\% and triple fraction of
10\% appropriate for Solar-mass stars \citep{Rag10,Duc13,Tok14b,Moe17}. Systems that are dynamically unstable at initialization are rejected. 

Our alternative models each differ from model STD in one aspect. In
model Q\_IN, we assume that the masses of the inner binary are uncorrelated,
as for the outer orbit of model STD. In model Q\_OUT, we make the
opposite assumption such that both the inner $q_{{\rm inner}}$ and
outer mass ratio $q_{{\rm outer}}\equiv m_{3}/(m_{1}+m_{2})$ are
distributed uniformly. In model A\_SANA, the distribution of the inner
and outer semi-major axes follow a power-law distribution $N\propto(logP)^{-0.55}$,
as observed in binaries with O- and B-type primaries \citep{San12}.
For A-type primaries in binaries, \citet{Riz13} found a log-normal
distribution of semi-major axes ($\mu=$0.95AU, $\sigma=1.35$), which
we adopt in model A\_RIZ with a maximum separation of $5\times10^{8}{\rm {R_{\odot}}}$.
Lastly, in model E\_CIRC we study the effect of eccentricities on
the collision rate. Initially these triples are circularized.

We consider stellar triples with $m_{1}>0.08{\rm {M_{\odot}}}$ and
$m_{2},m_{3}>0.008{\rm {M_{\odot}}}$. When drawing a mass from the
Kroupa IMF, we adopt a maximum mass of $100M_{\odot}$. To speed up
the simulations, we only simulate a subset of triples (comprising
a fraction $f_{{\rm param.space}}$ of parameter space, see Table
\ref{tbl:fraction}) that satisfy the following three requirements:
1) the inner binary can evolve into a double white dwarf within a
Hubble time, i.e. $0.95{\rm {M_{\odot}}}<m_{1},m_{2}<7.7{\rm {M_{\odot}}}$;
2) to avoid dissolution by the supernova $m_{3}<7.7{\rm {M_{\odot}}}$
; 3) to avoid mass transfer $a(1-e^{2})>2500{\rm {R_{\odot}}}$. 
We implicitly assume that the parameter space that we do not consider does not give rise to Ia-SNe. The
missing triples are taken into account in the normalization of the
rates.

\begin{table*}
\caption{{Distributions of the initial binary masses and orbital parameters
for the different models.}}
\begin{tabular}{|l|cccccc|}
\hline 
Model  & $q_{{\rm inner}}$ & $q_{{\rm outer}}$  & $a_{{\rm inner}}$ & $a_{{\rm outer}}$  & $e_{{\rm inner}}$ & $e_{{\rm outer}}$ \tabularnewline
\hline 
STD  & uniform  & uncorrelated  & log-uniform$^{1}$  & log-uniform  & thermal$^{2}$  & thermal \tabularnewline
Q\_IN  & \textbf{uncorrelated}  & uncorrelated  & log-uniform  & log-uniform  & thermal  & thermal \tabularnewline
Q\_OUT  & uniform  & \textbf{uniform}  & log-uniform  & log-uniform  & thermal  & thermal \tabularnewline
A\_SANA  & uniform  & uncorrelated  & \textbf{power-law}$^{3}$  & \textbf{power-law}  & thermal  & thermal \tabularnewline
A\_RIZ  & uniform  & uncorrelated  & \textbf{log-normal}$^{4}$  & \textbf{log-normal}  & thermal  & thermal \tabularnewline
E\_CIRC  & uniform  & uncorrelated  & log-uniform  & log-uniform  & \textbf{0}  & \textbf{0} \tabularnewline
\hline 
\end{tabular}

\label{tbl:init_param} $^{(1)}$ \citet{Abt83}; $^{(2)}$ \citet{Heg75};
$^{(3)}$ \citet{San12};$^{(4)}$ \citet{Riz13}
\end{table*}

\subsection{TRES\label{subsec:TRES}}

\texttt{TRES} is an astrophysical code to simulate the evolution
of stellar triples consistently \citep{Too16}; the code couples three-body
dynamics with stellar evolution including Kozai-Lidov oscillations,
tides, gravitational wave emission, and the effects of precession
and stellar winds. The dynamics is based on the secular approach up
to and including octupole-order (e.g. \citealt[for a review]{Nao16}).
Stellar evolution is simulated in a parametrized way through the
binary population synthesis code \texttt{SeBa} \citep{Por96,Too12}.
\texttt{TRES} is valid for simulating isolated coeval stellar triples.
Moreover, due to the usage of the secular approach, it is strictly
only appropriate to simulate the evolution of hierarchical triples. 

\texttt{TRES} is written in the Astrophysics Multipurpose Software
Environment, or \texttt{AMUSE} \citep{Por09,Por13}. It is a software
framework that includes codes from different astrophysical domains,
such as stellar dynamics, stellar evolution, hydrodynamics and radiative
transfer. \texttt{AMUSE} provides the user with a homogeneous interface
structure based on Python in which the community codes can be easily
used and coupled. \texttt{AMUSE} can be downloaded for free at
amusecode.org and github.com/amusecode/amuse. 

\subsection{Marginal hierarchical systems\label{subsec:Marginal-hierarchical-systems}}

We apply two methods to extract those triples that become marginally
hierarchical during their evolution. In method 1, we track the level
of hierarchy in the simulations with \texttt{TRES}. When a system
enters the ``quasi-secular'' regime, the simulation is stopped\footnote{Note that after a system enters the quasi-secular regime, it may still
take a long time before the collision occurs (see Eq. \ref{eq:tcol}).
If the triple is a 2WD with a relatively massive stellar tertiary,
stellar evolution may still play a role for the evolution of the system.
For all 2WDs in the quasi-secular regime $\apprle$10\% have a tertiary
star with mass above 0.95M$_{\odot}$. An exception to this is model
Q\_OUT, in which 63\% of quasi-secular 2WDs have a massive tertiary
that will evolve of the main-sequence in a Hubble time. The treatment
of the quasi-secular regime in this case is therefore not self-consistent.
A better modelling is beyond the scope of this project and could be
explored in the future. }, according to the following boundary condition we implemented in \texttt{TRES}:

\begin{equation}
\sqrt{1-e_{1}}<\sqrt{1-e_{{\rm crit}}}\equiv f_{{\rm crit}}\times5\pi\frac{m_{3}}{m_{1}+m_{2}}\left[\frac{a_{1}}{a_{2}(1-e_{2})}\right]^{3},\label{eq:e_crit}
\end{equation}
where $f_{{\rm crit}}$ is a numerical factor $\approx1$ (\citealt[ ]{Ant14};
see also \citealt{Ant12,Kat12}). If the inner eccentricity becomes
larger than $e_{{\rm crit}}$, the angular momentum of the inner orbit
can change by order of itself in one period. We note that with this
criterion, the SNIa progenitors in our models predominantly fulfil
the criteria \citep[See Eq. 7 in ][]{Kat12} to experience a clean
collision (see also \citealt{per+12} for a similar clean collision
criteria); all pericenter passages before the collision are large
enough such that tidal or general relativistic effects are negligible.
Cases where tidal interactions occur are treated by the regular secular
evolution coupled to tidal evolution.

In method 2, the simulations with \texttt{TRES} are performed until
time $t_{{\rm form}}$, which represents the time at which the triple
WD forms, or the double WD if the tertiary star does not evolve to
a WD in a Hubble time. 

We calculate analytically the maximum eccentricity $e_{{\rm max}}$
that the inner orbit can achieve afterwards based on the quadrupole
approximation for $e_{{\rm in}}\geq0$ \citep[e.g.][]{Kin07,Per09,Nao16}.
If $e_{{\rm max}}>e_{{\rm crit}}$ (Eq. \ref{eq:e_crit}), we assume
the system enters the quasi-secular regime. Such a simplified criteria
does not well capture the maximal eccentricity potentially reached
due to octupole-level perturbations, however the latter typically
become important for inner binaries with low-mass ratio, while WD-WD
binaries in our study always have a high mass ratio (typically above
0.85). Note that if a triple enters the quasi-secular regime before
$t_{{\rm form}}$, the evolution of the system is evaluated based
on the secular approach, which is not strictly valid in this regime.
However, it is likely that the system reaches high eccentricities
even within the secular approximation, such that Roche lobe overflow
develops. Mass transferring systems are not taken into account in
our rate estimates. Method 2 is less accurate than method 1, but allows
for more flexibility by varying $f$. 

\section{Results\label{sec:Res}}

\subsection{Formation}

\label{sec:form}

In order to calculate the rate of colliding WDs, we first study the
formation rate of a DWD with an outer companion, either another WD
(i.e. 3WD) or a low-mass stellar component (i.e. 2WD). In most of
our models, about 6\% of primordial triples evolve to a 3WD in a Hubble
time, and about 70\% to a 2WD (Table \ref{tbl:fraction}). An exception
to this is model Q\_OUT, in which the masses of the outer companion
are correlated to those of the inner binaries and the average tertiary
mass is therefore higher compared with the other models. In model
Q\_OUT 3WDs are formed more efficiently; about half of the primordial
triples evolve to a 3WD and a fifth evolve to become a 2WD. 

About 25\% of triples do not become a 2WD or 3WD (Table \ref{tbl:fraction}).
Mainly, these systems become dynamically unstable due to their stellar
and orbital evolution or the systems experience mass transfer. Regarding
the former, triples that become dynamically unstable due to mass loss
in the stellar winds are studied by \citet{per+12}.
This evolutionary channel can lead to a stellar collision involving
an (post-)AGB star. \citet{per+12} find that this channel could be
the dominant form of stellar collisions in the field. Regarding the
latter, in this paper we exclude triples that experience mass transfer.
Double WDs that form through mass transfer are expected to have short
orbits \citep[see e.g.][]{Too14}, which increases the hierarchy in
the triples and thereby do not evolve in the quasi-secular regime,
and are not likely to merge through collisions. Note that overall
the \textit{merger} rate between WDs is enhanced if the binary has
a tertiary companion, and some direct collisions may occur even outside
the quasi-secular regime. Indeed, non-quasi-secular mergers and collisions
in triples were studied by \citet{Ham13}, however, such mergers and
collisions do not contribute more than $10^{-3}$ of the Ia rate at
any given time. 

Next we consider those systems that become marginally hierarchical.
In method 1, we track the level of hierarchy at every timestep in
the simulations with \texttt{TRES}. We find that a few percent
of all simulated triples reach sufficiently high inner eccentricities
such that the systems enter the quasi-secular regime. The fraction
is highest in model Q\_OUT, which is related to the high average tertiary
masses in this model. Of interest here are the 2WDs and 3WDs that
become quasi-secular. All six models show that even though the formation
of a 2WD or 3WD is common (in the part of parameter space simulated
here), only a small percentage of the triples reach this state, i.e.
$0.2-0.5\%$. In method 2, we find similar percentages of $0.2-0.7\%$. 

\begin{table*}
\caption{Results of the population synthesis modelling. }
\label{tbl:fraction}

\begin{tabular}{|c|c|c|c|c|c|c|c|c|c|c|}
\hline 
\multirow{3}{*}{} & \multirow{3}{*}{method} & \multirow{3}{*}{$N_{{\rm sim}}$} & \multirow{3}{*}{$f_{{\rm param.space}}$} & \multicolumn{7}{c|}{fraction }\tabularnewline
\cline{5-11} 
 &  &  &  & mass  & dyn. & quasi- & \multirow{2}{*}{2WD} & 2WD quasi- & \multirow{2}{*}{3WD} & 3WD quasi- \tabularnewline
 &  &  &  &  transfer &  unstable & secular &  & secular &  & secular \tabularnewline
\hline 
STD & 1 & 50k & \multirow{2}{*}{$5.5\times10^{-3}$} & 0.093 & 0.11 & 0.018 & 0.70 & 0.0043 & 0.060 & -(\textless{}2e-5)\tabularnewline
 & 2 & 250k &  & 0.099 & 0.11 & - & 0.70 & 0.0050 & 0.060 & $1.3\times10^{-4}$\tabularnewline
\hline 
Q\_IN & 1 & 50k & \multirow{2}{*}{$9.7\times10^{-4}$} & 0.085 & 0.11 & 0.020 & 0.67 & 0.0039 & 0.060 & $2\times10^{-5}$\tabularnewline
 & 2 & 150k &  & 0.095 & 0.11 & - & 0.67 & 0.0050 & 0.062 & $1.3\times10^{-4}$\tabularnewline
\hline 
Q\_OUT & 1 & 50k & \multirow{2}{*}{$5.4\times10^{-3}$} & 0.10 & 0.095 & 0.046 & 0.19 & 0.0026 & 0.54 & $3.8\times10^{-4}$\tabularnewline
 & 2 & 100k &  & 0.12 & 0.099 & - & 0.19 & 0.0019 & 0.54 & $1.6\times10^{-3}$\tabularnewline
\hline 
A\_SANA & 1 & 50k & \multirow{2}{*}{$1.8\times10^{-3}$} & 0.11 & 0.11 & 0.015 & 0.68 & 0.0030 & 0.061 & $4\times10^{-5}$\tabularnewline
 & 2 & 150k &  & 0.12 & 0.11 & - & 0.68 & 0.0043 & 0.062 & $9.4\times10^{-5}$\tabularnewline
\hline 
A\_RIZ & 1 & 50k & \multirow{2}{*}{$3.0\times10^{-3}$} & 0.14 & 0.12 & 0.012 & 0.64 & 0.0022 & 0.056 & -\tabularnewline
 & 2 & 100k &  & 0.14 & 0.12 & - & 0.65 & 0.0028 & 0.055 & $3.0\times10^{-5}$\tabularnewline
\hline 
E\_CIRC & 1 & 50k & \multirow{2}{*}{$1.0\times10^{-2}$} & 0.062 & 0.10 & 0.019 & 0.72 & 0.0045 & 0.067 & -\tabularnewline
 & 2 & 15k &  & 0.067 & 0.099 & - & 0.73 & 0.0069 & 0.068 & $2.8\times10^{-4}$\tabularnewline
\hline 
\end{tabular}
\end{table*}

\subsection{SNe rates and delay time distributions\label{subsec:col}}

We find that head-on collisions between carbon-oxygen WDs from wide
isolated triples happen at a rate of a few times $10^{-7}$ per Solar
mass of created stars. The rates of the different models and methods
are given in Table \ref{tbl:rates}. In comparison, the observed rate
of supernova Type Ia in field galaxies is about $10^{-3}\,{\rm M_{\odot}^{-1}}$
\citep[e.g.][]{Mao+14,Mao17} and therefore the contribution the isolated-triples
channel to the SNIa rate is of the order of $0.01-0.1\%$.

The different models of primordial triples give rise to up to an order
of magnitude uncertainty in the synthetic Ia-SNe rates. The highest
rates are expected if the inner and outer orbits are circularized
(model E\_CIRC). In this case fewer systems will undergo mass transfer,
as indicated in Table \ref{tbl:fraction} by the large fraction of
parameter space that is simulated $f_{{\rm param.space}}$, and the
small fraction of systems experiencing mass transfer in the simulated
triples. Consequently, these systems will follow a different evolutionary
channel than considered here. The lowest rates are anticipated if
the masses of the three stars are not correlated to one another (model
Q\_IN). In this case the average mass of the secondary is low, its
evolutionary timescale is long, such that fewer triples will harbour
two WDs in the inner binary (see $f_{{\rm param.space}}$ in Table
\ref{tbl:fraction}). Furthermore, as the average mass of the tertiary
is low, the dynamical effect of the tertiary on the inner binary
is smaller, and fewer systems enter the quasi-secular regime. 

Another aspect that affects the predicted collision rate is the extent
of the quasi-secular regime. So far we have adopted a sharp boundary
between the secular and quasi-secular regime, i.e. $f_{crit}=1$. In
reality, the reliability of the secular approximation deteriorates
gradually when approaching the critical boundary. Assuming that the
secular approximation falls short at $f_{crit}=2$ ($f_{crit}=10$
) and using method 2, the collision rate increases by about a factor
$\sim2-3\,(\sim10-40).$

The collision rate depends also on the abundance of triples, i.e. the
triple fraction. Here we have assumed a triple fraction of 10\% and
a binary fraction of 40\%. These values are based on observation of
Solar-type stars \citep{Rag10,Duc13,Tok08,Tok14b}. However, several
studies have shown that the binary fraction varies with the stellar
type of the primary \citep{Rag10,Duc13,Kle+17,Moe17}. The binary
fraction increases with primary mass, and there are indications that
the triple fraction follows a similar trend \citep{Rem11,San14,Moe17}.
Assuming a triple fraction of 25\% and a binary fraction of 60\% (appropriate
for A-type stars), the collision rates in Table \ref{tbl:rates} increase
by a factor $\sim2$. Taking the most optimistic and likely not realistic
assumptions, i.e. a high triple fraction, $f_{crit}=10$ and model
E\_CIRC one can reach a level of $\sim6$\% of Ia-SNe from this channel.
A more plausible fraction would of the order of $0.1-1\%$ of Ia-SNe.

\begin{table}
\caption{{Time-integrated collision rate of CO-CO WDs per Solar mass of created
stars. The different models are described in Sect.\ref{subsec:Primordial-triples},
and the methods in Sect.\ref{sec:met}.}}
\begin{tabular}{|l|ll|}
\hline 
Model & \multicolumn{1}{l}{Method 1} & Method 2\tabularnewline
\hline 
STD  & 3.0 $\times$ 10$^{-7}$ & 4.7 $\times$ 10$^{-7}$\tabularnewline
Q\_IN  & 4.7 $\times$10$^{-8}$  & 7.5$\times$10$^{-8}$\tabularnewline
Q\_OUT  & 1.5$\times$10$^{-7}$ & 2.6 $\times$ 10$^{-7}$\tabularnewline
A\_SANA  & 1.0$\times$10$^{-7}$ & 2.0 $\times$ 10$^{-7}$\tabularnewline
A\_RIZ  & 2.4$\times$10$^{-7}$  & 4.3 $\times$ 10$^{-7}$ \tabularnewline
E\_CIRC  & 6.1$\times$ 10$^{-7}$ & 1.7 $\times$10$^{-6}$\tabularnewline
\hline 
 Observed$^{1}$ & \multicolumn{2}{l|}{$(1.3\pm0.1)\times10^{-3}$, $(1.6\pm0.3)\times10^{-3}$}\tabularnewline
\hline 
\end{tabular}
\begin{flushleft}
$^{(1)}$\citet{Mao17,Mao+14}
\end{flushleft}
\label{tbl:rates}
\end{table}

\begin{figure}[h!]
\centering{}\includegraphics[width=1\columnwidth]{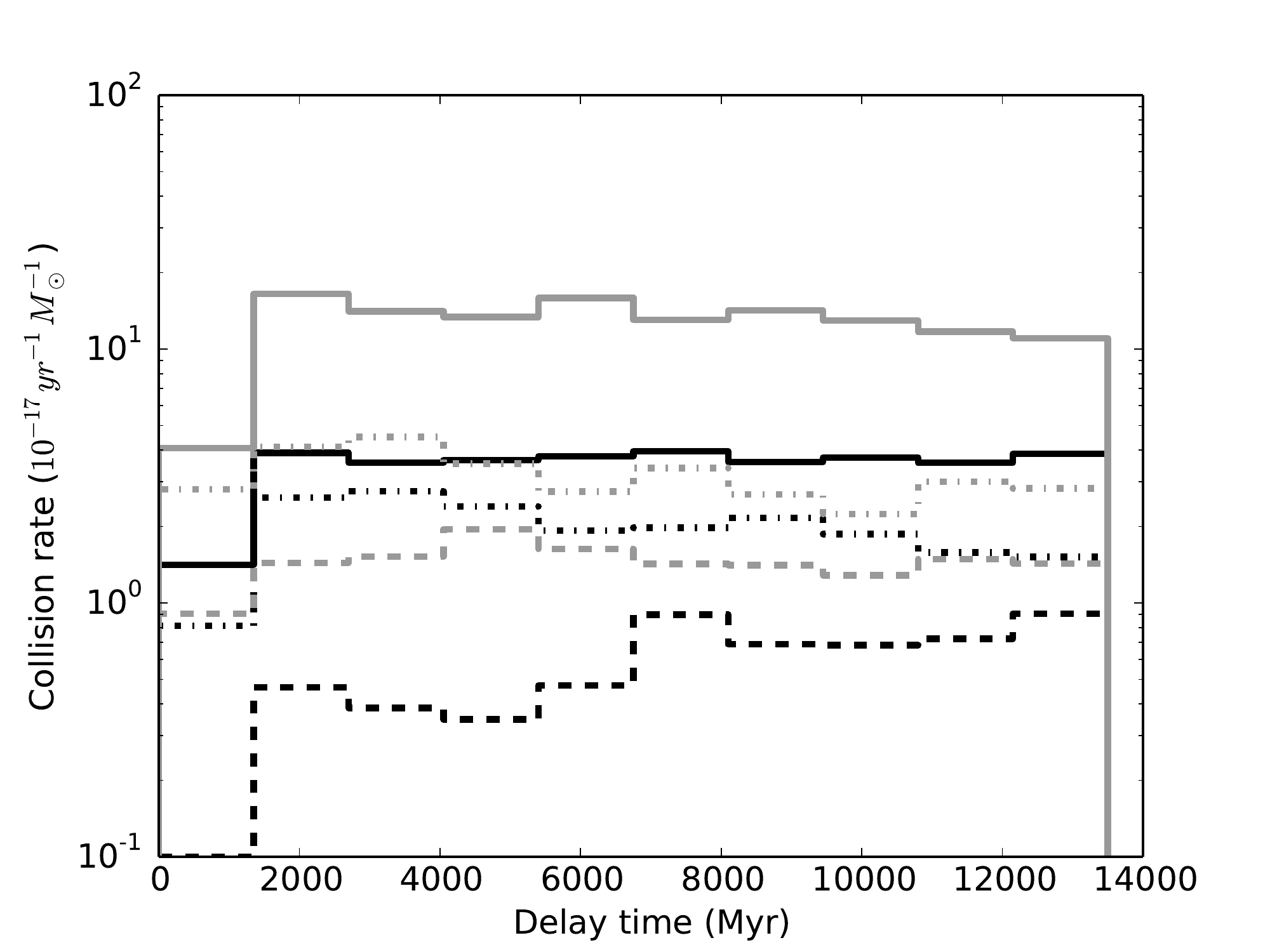} \caption{Delay time distribution of head-on collisions between white dwarfs
in isolated triples. The rate of collisions is given per $10^{17}$yr
per Solar mass of created stars. The different line-styles correspond
to the different models in method 2. The DTD is approximately uniform
in time. Both the normalization and the shape of the DTD is in clear
contradiction with observations. \label{fig:DTD} }
\end{figure}

\begin{figure}[h!]
\centering{}\includegraphics[width=1\columnwidth]{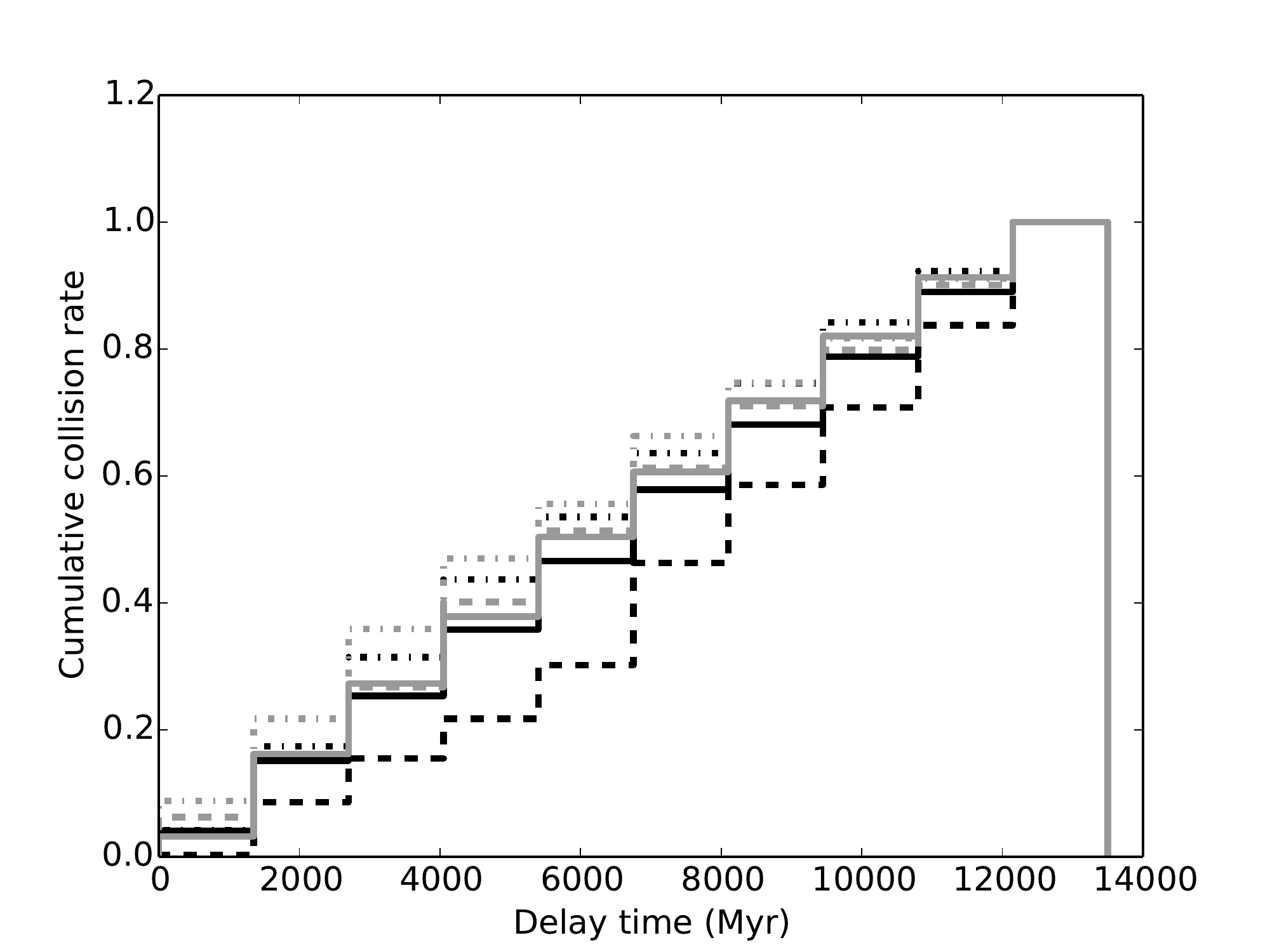}
\caption{Cumulative delay time distribution of head-on collisions between white
dwarfs in isolated triples. The collision rates are normalized to
1. \label{fig:DTD_cum} }
\end{figure}

The collision rate as a function of time since a single burst of star
formation, i.e. the delay time distribution (DTD) is shown in Fig.
\ref{fig:DTD} and \ref{fig:DTD_cum}. These shown DTDs are for all
models in method 2. The predicted DTDs from method 1 (not shown) are
in good agreement with those shown in the figures. We find that the
DTDs of all models and methods have a very distinct shape, namely
the delay times are distributed uniformly in time. On the other hand,
the observed SNIa DTD decreases strongly with time. The characteristic
shape of the observed DTD is $dN/dt\propto t^{\beta}$ with $\beta\thickapprox-1$
\citep{Gra11,Gra13,Her17}. 

Due to the shape of the predicted and observed DTD, the largest contribution
from head-on collisions in isolated triples is expected at long delay
times. At these times, the observed SNIa rate in field galaxies is
about $10^{-14}{\rm yr^{-1}}{\rm M_{\odot}^{-1}}$ (Table \ref{tbl:DTD}),
and our maximum predicted collision rate is about $10^{-16}{\rm yr^{-1}}{\rm M_{\odot}^{-1}}$
(model E\_CIRC). At best, collisions in triples contribute about 1\%
to the SNIa rate at late times.

\begin{table}

\caption{The delay time distribution at long delay times in field galaxies.
Taken from \citet{Mao17}.\label{tbl:DTD}}
\begin{tabular}{|c|c|c|}
\hline 
Delay (Gyr) & DTD & Reference\tabularnewline
\hline 
\hline 
$8.1_{-5.7}^{+5.7}$ & $3_{-0.6}^{+1.5}$ & \citet{Mao11}\tabularnewline
\hline 
$8.1_{-5.7}^{+5.7}$ & $1.8_{-0.4}^{+0.4}$ & \citet{Mao12}\tabularnewline
\hline 
$8.1_{-5.7}^{+5.7}$ & $4.5_{-0.6,-0.5}^{+0.6,+0.3}$ & \citet{Gra13}\tabularnewline
\hline 
\end{tabular}
\end{table}

\subsection{Masses and mass-ratios of colliding WDs}

\begin{figure}[h!]
\centering{}\includegraphics[width=1\columnwidth]{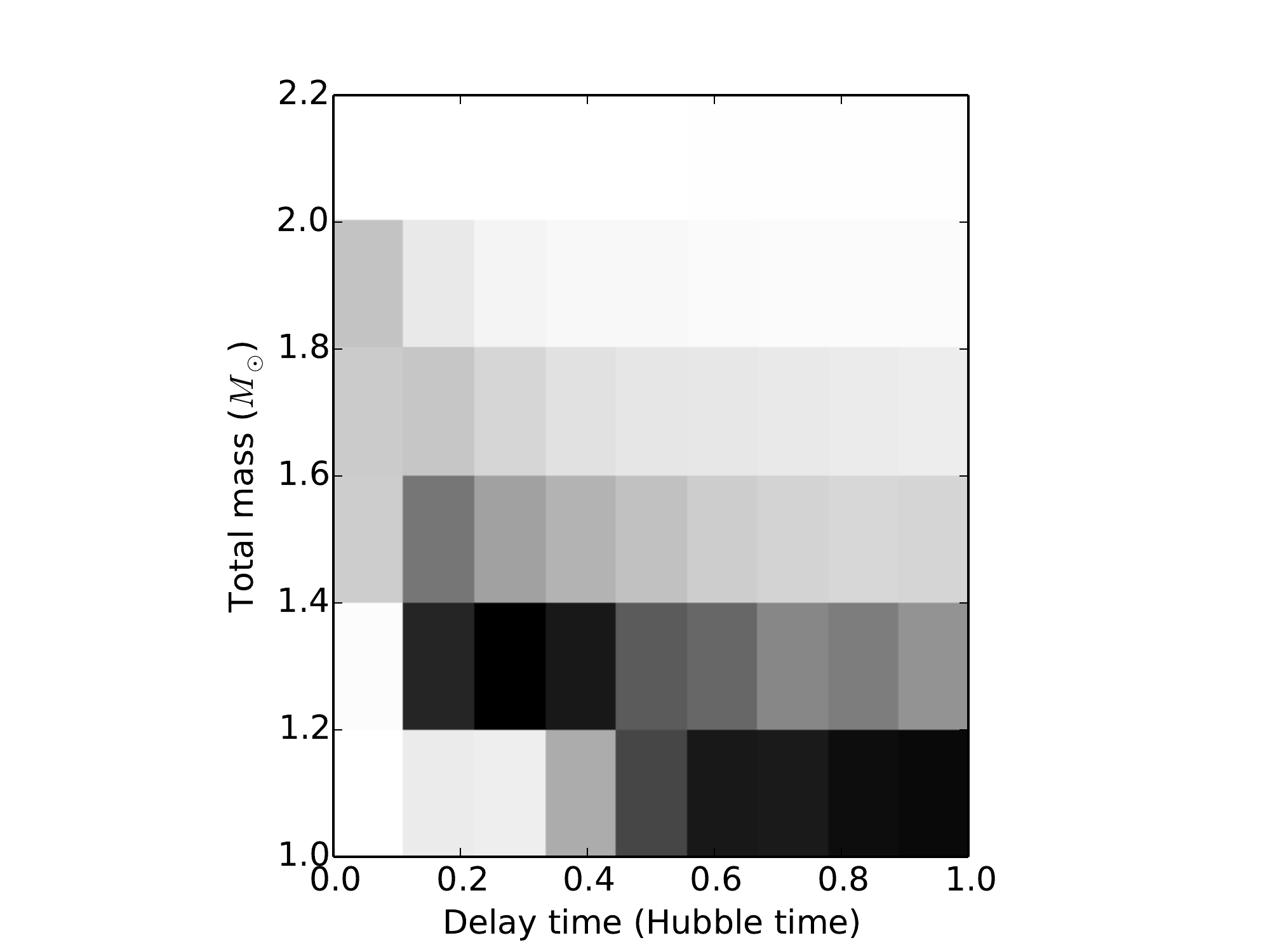} \caption{\label{fig:Mtot} Total mass of the colliding white dwarfs of the
inner binary as a function of the delay time for model STD and method
2. Delay time is given as a fraction of the Hubble time, here taken
as 13.5 Gyr. The grey scale is a density of objects on a linear scale.}
\end{figure}

\begin{figure}[h!]
\centering{}\includegraphics[width=1\columnwidth]{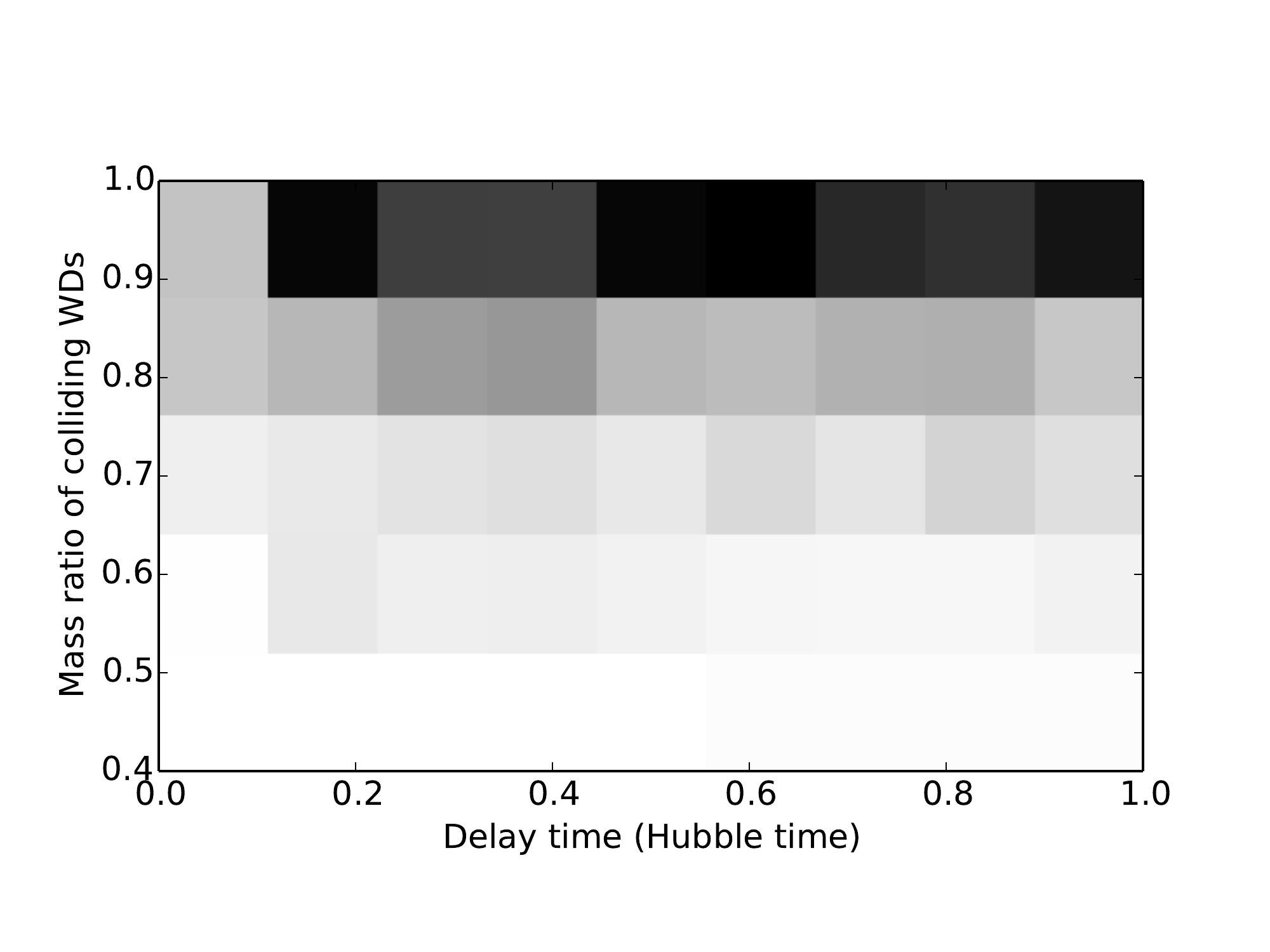} \caption{\label{fig:q} Mass ratio of the colliding white dwarfs of the inner
binary as a function of the delay time for model STD and method 2.
Delay time is given as a fraction of the Hubble time, here taken as
13.5 Gyr. The grey scale is a density of objects on a linear scale.}
\end{figure}

Now we turn to the masses of the colliding WDs. The combined mass
of the colliding WDs and their mass ratios are shown in Fig.\,\ref{fig:Mtot}~and~\ref{fig:q}.
These figures represent model STD using method 2, but all models show
similar behaviours. The stellar evolution timescales are long for (single)
low-mass stars i.e. low-mass WD progenitors. Therefore WD-WD systems
with high total mass tend to form earlier, and therefore collide following
a short delay time, while low-mass systems form later and thereby
give rise to long delay times. 

The individual masses of the colliding WDs are close to one another
(Fig.\ref{fig:q}). This is expected as the initial-to-final mass
relation for WDs is fairly flat, e.g. \citet{Kal08} who find $M_{{\rm final}}=(0.109\pm0.007)M_{{\rm initial}}(M_{\odot})+(0.394\pm0.025)M_{\odot}$. 

\label{sec:charac}

The masses of the colliding WDs are important for the amount of $^{56}$Ni that can be synthesized in the collision \citep[see e.g.][]{Gar13}. 
Early works of hydrodynamical simulations of colliding WDs have found a minimum mass ratio ($q \gtrsim 0.6-0.7$) and/or total mass $\gtrsim 1.0-1.2{\rm {M_{\odot}}}$ necessary for an explosion to take place. On the other hand, at higher resolution \citet{Kus13} found all their collisions to produce enough $^{56}$Ni in order to appear as a SNIa. 
In our simulations, the mass ratios are typically $\sim 0.9-1$ (Fig.\ref{fig:q}), such that a minimum mass ratio of $\sim 0.6-0.7$ does not effect the predicted collision rate significantly. On the other hand if the minimum total mass of the colliding WDs to produce a SNIa-like event is 1.2M$_{\odot}$, the synthetic rates given in this paper can be seen as an upper limit (Fig.\ref{fig:Mtot}).

\section{Discussion and Summary}

\label{sec:concl}

In this paper we studied the rate of WD-WD direct collisions induced
by secular and quasi-secular evolution in triple stellar systems,
and their potential to explain the origin of typical type Ia SNe.
Though triple secular evolution was shown to produce only low-rates
of direct WD-WD collisions \citep{Ham13}, it was suggested that
less hierarchical triple systems can evolve through \emph{quasi-}secular
evolution \citep{Ant12} leading to high rates of WD-WD collisions
in triple \citep{Kat12}. In order to study this possibility we used
a novel triple population synthesis models \texttt{TRES}, coupled
to simplistic application of quasi-secular evolution, when relevant,
and explored the properties of such triple-formed SNe, including their
rates, total mass of the colliding WDs, the WDs mass-ratios and the
expected delay time distribution. Given the many uncertainties in
the properties and quasi-secular evolution of triple systems, we constructed
a range of plausible models for the triple progenitor population,
and employed several simplified models to account for the quasi-secular
evolution involved. Though the predicted rates may range over an order
of magnitude, depending on the chosen model, all models predict no
more than $\sim0.1$\% of regular type Ia may arise from WD collisions
in isolated triples. Moreover, the delay time distribution of SNe
from this channel is distributed uniformly over time, and is therefore
inconsistent with that inferred from observations ($\sim t^{-1}$). 

Many of the potential progenitors that initially have high inclinations
and relatively weak hierarchy, i.e. ``active'' triples susceptible
to the quasi-secular evolution already dynamically evolve into mass-transfer,
mergers or collisions during the main-sequence of giant-branch stages,
and never produce WD-WD binaries for which collisions can be induced by
the third stellar companion. Possible channels to introduce more ``active''
triples with WD-WD inner binaries in the relevant phase space could
be through the perturbations of non-''active'' triple, such as a
triple with low-mutual inclination which would not quasi-secularly
evolve significantly otherwise. In stellar clusters triples could
be perturbed by other stars in the cluster and thereby change their
orbital parameters. The evolution of such non-isolated perturbed triples
is not considered here, however; the total number of triples in clusters
is relatively small, both due to the total number of stars in clusters
in general in addition to the small fraction of triples, which would
need to be sufficiently compact (``hard'') as not to be disrupted
by encounters with other stars. We therefore do not expect triples
in cluster to contribute significantly to the formation of type Ia
SNe.

Triples in the field might also be susceptible to flyby encounters
by field stars. However, such flybys introduce negligible changes
in the triple orbits, and at most minor changes in the orbits of the
widest triples. Study of flybys in very wide triples will be explored
elsewhere, but these too are not expected to contribute significantly.
We conclude that triples, and in particular isolated field triples,
are likely to produce only a small fraction (at most a percent) of
type Ia SNe. 

\bibliographystyle{aasjournal}
\bibliography{sne,bibtex_silvia_toonen}

\acknowledgements{ST gratefully acknowledges support from the Netherlands Research Council NWO
(grant VENI {[}nr. 639.041.645{]}). 
HBP and ST gratefully acknowledge support
from the Israel science foundation I-CORE program 1829/12.
ASH gratefully acknowledges support from the Institute for Advanced Study, and from NASA grant NNX14AM24G.}
\end{document}